\documentclass[twocolumn]{aastex63}
\NewPageAfterKeywords
\usepackage{lipsum}
\usepackage{microtype}
\usepackage[many]{tcolorbox}

\definecolor{myred}{HTML}{FB6542}
\definecolor{dodgerblue}{RGB}{30, 144, 255}
\definecolor{myblue}{HTML}{375E97}
\definecolor{tableblue}{HTML}{1A73C9}
\definecolor{mygrey}{HTML}{363636}

\newtcolorbox{bangbox}[2][]
{
enhanced,
  colframe=white,
  title={#2},
  width=6.2in, 
  colback=tableblue,
  coltext=white,
  center title,
  colbacktitle=myred,
  coltitle=white,
  fonttitle=\bfseries
}

\newcommand{\lasp}{
	Laboratory for Atmospheric and Space Physics,
	University of Colorado,
	3665 Discovery Dr.,
	Boulder, CO 80303-7814, USA
	}

 \newcommand{\cuaero}{ 
    Aerospace Engineering Sciences,
	University of Colorado,
    3775 Discovery Dr.,
	Boulder, CO 80309-0390, USA
	}
 
\newcommand{\jhuapl}{
	Applied Physics Laboratory,
	Johns Hopkins University,
	11100 Johns Hopkins Rd.,
	Laurel, MD 20723, USA
	}
\newcommand{\gsfc}{
	NASA Goddard Space Flight Center,
	8800 Greenbelt Rd.,
	Greenbelt, MD 20771, USA
	}
\newcommand{\msfc}{
	NASA Marshall Space Flight Center,
	Martin Rd. SW,
	Huntsville, AL 35808, USA
	}
\newcommand{\swri}{
	Southwest Research Institute,
	1050 Walnut St., Suite 300,
	Boulder, CO 80302, USA
	}
\newcommand{\hao}{
	High Altitude Observatory,
    3080 Center Green Dr.,
	Boulder, CO 80301, USA
	}
\newcommand{\nwra}{
	Northwest Research Associates,
    3380 Mitchell Ln.,
	Boulder, CO 80301, USA
	}
\newcommand{\iowa}{
	University of Iowa,
    203 Van Allen Hall,
	Iowa City, IA 52242-1479, USA
	}
\newcommand{\cfa}{
	Center for Astrophysics $\vert$ Harvard Smithsonian,
    60 Garden St.,
	Cambridge, MA 02138, USA
	}
\newcommand{\uta}{
	University of Texas at Arlington,
    701 S. Nedderman Drive,
    Arlington, TX 76019, USA
	}
\newcommand{\unh}{
	University of New Hampshire,
    105 Main St,
    Durham, NH 03824, USA
	}
\newcommand{\bwx}{
	BWX Technologies,
    800 Main St,
    Lynchburg, VA 24504, USA
	}

 \newcommand{\predsci}{
    Predictive Science Inc.,
    9990 Mesa Rim Rd, Suite 170,
    San Diego, CA 92121, USA
    }

\begin{document}

\title{\large Small Platforms, High Return: 
The Need to Enhance Investment in Small Satellites for Focused Science, Career Development, and Improved Equity}

\author[0000-0002-3783-5509]{James Paul Mason}
\affiliation{\jhuapl}

\author[0000-0001-9091-7768]{Robert G. Begbie}
\affiliation{\cuaero}

\author[0000-0001-5024-014X]{Maitland Bowen}
\affiliation{\lasp}

\author[0000-0001-8702-8273]{Amir Caspi}
\affiliation{\swri}

\author[0000-0003-4372-7405]{Phillip C. Chamberlin}
\affiliation{\lasp}

\author[0000-0001-8137-6357]{Amal Chandran}
\affiliation{\lasp}

\author[0000-0002-9163-6009]{Ian Cohen}
\affiliation{\jhuapl}

\author[0000-0001-7416-2895]{Edward E. DeLuca}
\affiliation{\cfa}

\author[0000-0002-5084-4661]{Alfred G. de Wijn}
\affiliation{\hao}

\author[0000-0001-5661-9759]{Karin Dissauer}
\affiliation{\nwra}

\author[0000-0001-7143-2730]{Francis Eparvier}
\affiliation{\lasp}

\author[0000-0002-5497-7867]{Rachael Filwett}
\affiliation{\iowa}

\author[0000-0001-9831-2640]{Sarah Gibson}
\affiliation{\hao}

\author[0000-0003-0021-9056]{Chris~R.~Gilly} 
\affiliation{\lasp}

\author[0000-0001-9139-8939]{Vicki Herde}
\affiliation{\lasp}

\author[0000-0003-1093-2066]{George Ho}
\affiliation{\jhuapl}

\author[0000-0001-9200-9878]{George Hospodarsky}
\affiliation{\iowa}

\author[0000-0002-1470-4266]{Allison Jaynes}
\affiliation{\iowa}

\author[0000-0001-5533-5498]{Andrew R. Jones}
\affiliation{\lasp}

\author[0000-0002-7077-930X]{Justin C. Kasper}
\affiliation{\bwx}

\author[0000-0001-8163-1696]{Rick Kohnert}
\affiliation{\lasp}

\author[0000-0002-2992-558X]{Zoe Lee}
\affiliation{\lasp}

\author[0000-0002-8767-7182]{E. I. Mason}
\affiliation{\predsci}

\author[0000-0001-9751-6481]{Aimee Merkel}
\affiliation{\lasp}

\author[0000-0003-4681-8679]{Rafael Mesquita}
\affiliation{\jhuapl}

\author[0000-0002-4103-6101]{Christopher S. Moore}
\affiliation{\cfa}

\author[0000-0002-8608-2822]{Romina Nikoukar}
\affiliation{\jhuapl}

\author[0000-0002-8306-2500]{W. Dean Pesnell}
\affiliation{\gsfc}

\author[0000-0002-7628-1510]{Leonardo Regoli}
\affiliation{\jhuapl}

\author[0000-0002-6172-0517]{Sabrina Savage}
\affiliation{\msfc}

\author[0000-0002-0494-2025]{Daniel B. Seaton}
\affiliation{\swri}

\author[0000-0002-2526-2205]{Harlan Spence}
\affiliation{\unh}

\author[0000-0002-5305-9466]{Ed Thiemann}
\affiliation{\lasp}

\author[0000-0002-7407-6740]{Juliana T. Vievering}
\affiliation{\jhuapl}

\author[0000-0002-2463-4716]{Frederick Wilder}
\affiliation{\uta}

\author[0000-0002-2308-6797]{Thomas N. Woods}
\affiliation{\lasp}

\begin{abstract}
\begin{bangbox}{Key Points}
    \begin{itemize}
        \item Small satellites presently fill an appropriate niche in the cost versus mission-class/risk ecosystem
        \item Small satellites are highly productive as measured by papers/year/\$M
        \item Small satellites present the all-too-rare opportunity to further develop a career in spaceflight leadership
        \item Equity goals can be met -- and met soon -- by increasing the number of opportunities to lead small satellite missions
        \item Present small satellite funding levels are a bottleneck and prioritizing investment in them will enhance the returns described above
    \end{itemize}
\end{bangbox}
\vspace{1cm}
\end{abstract}
\setcounter{page}{0}

\section{Executive Summary}
In the next decade, there is an opportunity for very high return on investment of relatively small budgets by elevating the priority of smallsat funding in heliophysics. We've learned in the past decade that these missions perform exceptionally well by traditional metrics, e.g., papers/year/\$M \citep{Spence2022}. It is also well established that there is a ``leaky pipeline" resulting in too little diversity in leadership positions (\href{https://www.nationalacademies.org/our-work/increasing-diversity-in-the-leadership-of-competed-space-missions}{National Academies Report}). Prioritizing smallsat funding would significantly increase the number of opportunities for new leaders to learn -- a crucial patch for the pipeline and an essential phase of career development. \textbf{At present, however, there are far more proposers than the available funding can support, leading to selection ratios that can be as low as 6\%} -- in the bottom 0.5th percentile of selection ratios across the history of ROSES\footnote{NASA HFORT received 16 compliant proposals in the most recent call at the time of writing (HFORT2020). HFORT2022 expects to make only 1-2 selections. The number of proposals is not likely to decrease with time given historical trends, meaning that we may soon be facing selection selection ratios no higher than 6-12\%. In the \href{https://science.nasa.gov/researchers/sara/grant-stats}{list of 1180 ROSES selections made since 2003}, a 6\% selection ratio is in the bottom 0.5th percentile}. \textbf{Prioritizing SmallSat funding and substantially increasing that selection ratio are the fundamental recommendations being made by this white paper.}

\begin{figure*}
    \centering
    \includegraphics[width=.9\textwidth]{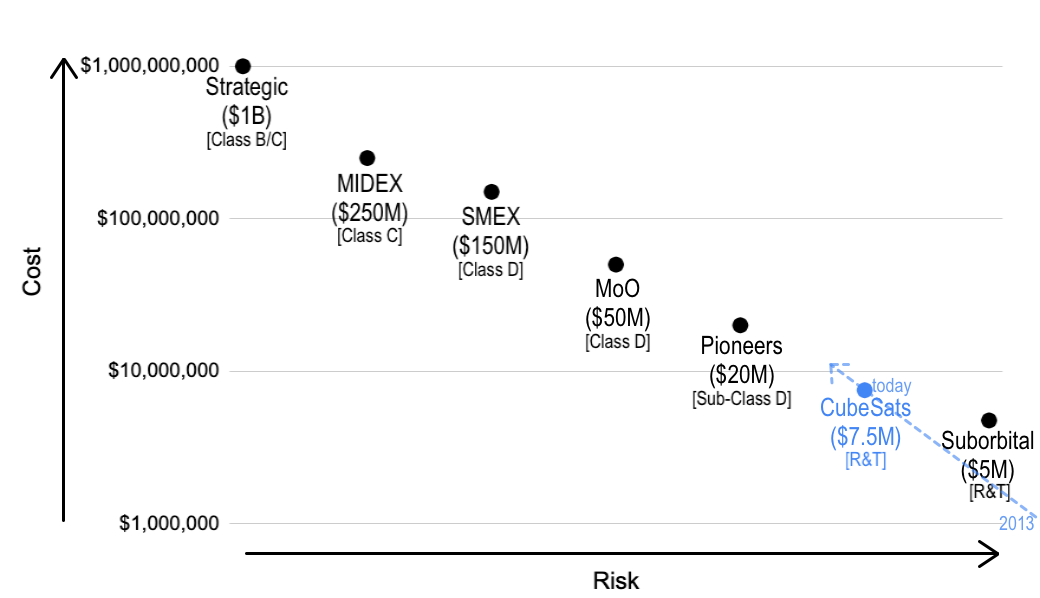}
    \caption{Typical cost of missions versus an arbitrary technical risk scaling guided by mission class. In the past decade, CubeSats moved up from $\sim$\$1M/mission and settled into a very natural position in the overall portfolio (dashed blue line), but the trend toward less risk tolerance and higher cost should stop now. Note that the vertical axis is log scaled. ``R\&T" is "Research and Technology", as defined in NASA Procedural Requirements 7120.8A.}
    \label{fig:ecosystem}
\end{figure*} 

Figure \ref{fig:ecosystem} illustrates the current mission ``ecosystem" as a function of cost and risk. Over the past decade, SmallSats (CubeSats as well as the new astrophysics Pioneers class) have come to fill the large gap between the suborbital class and Missions of Opportunity (MoO). In order to increase the SmallSat selection rate, their costs -- which are driven, in part, by how much risk is allowed -- should be stabilized early in the next decade. If the costs rise at the same rate that the program element total budgets do, the selection ratio will continue to suffer.

\section{High Return on Investment (ROI)}
\label{sec:roi}
We categorize the returns received into 1) focused science, 2) career development, and 3) improved equity. There is interplay between these categories that will become apparent.

\subsection{Focused Science}
\label{sec:science}
There are niche scientific problems that are \textit{best} addressed with small satellites; larger missions are simply unnecessary in these cases. Reports, such as ``Achieving Science with CubeSats: Thinking Inside the Box" \citep{NRC2016} suggest exactly this. The scope of the proposed scientific question(s) and the scope of the mission should be congruent. Narrower scope, however, does not imply a lack of scientific importance. For example, measurements of high energy particles in Earth's radiation belts made by the CSSWE CubeSat contributed to a large enough impact on our understanding of the magnetosphere to warrant two publications in Nature \citep{Baker2014, Li2017}. In order to emphasize this point, \citet{Spence2022} compared a representative set of missions spanning mission class: flagship/strategic, MIDEX, and SMEX; versus CubeSats (mostly NSF-funded, $\sim$1M education-oriented but science-motivated missions). While imperfect, an easy and commonly-used metric for scientific productivity is the number of resultant papers. Of course, the CubeSats had far fewer average total papers (approximately 1\% the average of large missions). However, when normalized by year since launch and mission cost, the \textbf{CubeSats performed four times better}. There are many reasons for this, including launch costs, risk posture, and mission complexity. There are important measurements that require large platforms; we are emphasizing that both large \textit{and} small missions have value. The important point here is that any lingering narrative that small satellites are not worth the investment should be retired. Even based on this traditional metric alone -- which only accounts for scientific productivity -- small satellites can have an exceptional return on investment (ROI). 

Additionally, the miniaturization of spacecraft bus technologies has created an incentive to also miniaturize instrumentation in order to take advantage of these new platforms. Many CubeSats can be cited as examples of this but here we will only provide the example of the miniaturization of the Van Allen Probes (VAP) Relativistic Electron-Proton Telescope (REPT) for the Colorado Student Space Weather Experiment (CSSWE) CubeSat REPT integrated little experiment (REPTile); these missions flew in highly complementary orbits (polar and equatorial) that enhanced the science output for both missions. Looking forward, solar sails are a bus technology in active development presently that will continue to drive demand for very low mass instruments. Moreover, the abundance of future launches to the moon and beyond open up rideshare opportunities that have been rare to date; this too will increase the demand for small, standardized payloads that can address focused science questions from comparatively rare orbits \citep{NAS2020, ASU2022}.

However, miniaturization does have some limits. Some focused science measurements require slightly more than the typical \$7.5M but still substantially less than a \$50M MoO. For example, the 3U CubeSat, LightSail-2, was dedicated to and successful in testing a solar sail \citep{Spencer2020}; but employing this technology to obtain heliophysics measurements beyond Earth orbit would require a higher than typical budget for (at least some) radiation hard electronics, additional associated testing, a longer mission duration to account for the cruise phase to the target destination(s), and/or communications via the Deep Space Network. Another example is constellations, which are a major benefit of these small platforms. It is just possible to fund the smallest possible constellation -- 2 satellites -- for \$7.5M. Over the past decade, NASA Heliophysics has been selecting an increasing number of constellations in the higher mission classes (e.g., VAP, MMS, PUNCH, TRACERS, EZIE) -- a clear indicator of the compelling science that can only be done with a constellation. Today, this benefit is largely restricted to mission classes at or above MoOs. These examples illustrate the need for a mission class to fill in the wide gap between CubeSats and MoOs (Figure \ref{fig:ecosystem}). NASA Astrophysics has recently created such a class -- Pioneers -- with a per-mission cost cap of \$20M. Heliophysics Pioneers would enable new science while remaining less than half the cost of MoOs, the next class up.

Finally, we have evidence that there are more excellent ideas being proposed by the community than existing budgets can support. Due to budget restrictions, HFORT2020 was forced to decline 50\% of their top tier proposals, all of which were rated ``Excellent" by the panel. As underlying spacecraft technology continues to improve and as instruments continue to miniaturize, new, comprehensive approaches will emerge for addressing the science topics to be recommended by the Decadal Survey. It is clear that more funding for small satellites would be a wise investment. For example, a modest enhancement equivalent to a single MIDEX mission could fund $\sim$30 additional small satellite missions. 

\subsection{Career Development}
\label{sec:career}
Large missions benefit from leaders that already have leadership experience, even if the scope of that previous experience was smaller. Such leaders already have refined skills in managing disparate teams, setting up systems to ensure clear lines of communication everywhere they are needed, appropriate delegation, conflict resolution, and managing potential risks for the mission. It benefits everyone to do this refinement -- this learning -- when the stakes are lower. Mistakes tend to be less costly but the lessons learned just as memorable. 

Moreover, the experience of working on a small mission itself opens up opportunities to work on larger missions that likely would not have occurred otherwise. As with anything, expertise tends to stack. Gaining recognition in the community through conference presentations and papers about various aspects of the mission and its science means that when new mission concepts are percolating among a group of people, it's more likely that person's name will be readily accessible to lead an important aspect of the mission (an instrument on a MIDEX, for example). This networking is effective in the other direction as well: everyone on a small satellite team has the opportunity to interact with engineers, instrument scientists, project managers, etc. -- people with whom they likely would not have interacted before. People in leadership positions also build deeper relationships with financial managers and administrators than is required for science research and analysis. All of this means that the person will already ``speak the language" and have an existing network of well-established relationships outside their own scientific discipline to support them if they find themselves leading a larger mission.

Some groups have informally implemented a ``see one, do one, teach one" practice for small missions such as repeated launches of a sounding rocket over the course of a few years. Often times a graduate student or postdoc will participate in the rocket payload preparation and launch but have a relatively minor set of responsibilities -- their primary job is to learn. On the subsequent launch, they lead important parts of the project and are deep into many of the hands-on aspects. They may go on to propose their own sounding rockets or small satellite missions from there, but there is an expectation that on the next launch, they will help train the newest members of the team. Such a practice is beneficial for everyone involved and ensures there are always experienced hands around to mitigate risk. If such a practice could be formalized and be done cross-institutionally for small satellites -- essentially a SmallSat Academy -- the benefits could reach a wider audience.

Again, prioritizing funding in small satellite programs would expand all of these benefits to \textit{more people}. 

\subsection{Improved Equity}
\label{sec:equity}
It is clear that improving Diversity, Equity, and Inclusion (DEI) is a high priority for all of agencies that the Decadal Survey targets, from the top of those organizations to the bottom. For example, it is explicitly called for in NASA's ``Vision for Scientific Excellence" and it is becoming actionable in places like the new requirement for some ROSES proposals to include an explicit ``Inclusion Plan." An extremely relevant report recently commissioned by NASA was released: ``Increasing Diversity, Equity, Inclusion, and Accessibility in the Leadership of Competed Space Missions" \citep{NAS2022}. Everything in that report is relevant to the present discussion but even a summary is beyond the scope of this white paper. Instead, we will briefly discuss the specific interplay of this topic with the previous section. 

The salient issue is commonly referred to as the ``leaky pipeline." At each step from elementary school through graduate school to early career and then senior professional, there is attrition that is not proportional to the underlying population demographics. The focus of this white paper is primarily at the high school to early-career segment of the pipeline; this is where bolstering the investment in small satellites can provide a patch for part of the leak. The cost of each new small satellite mission is small. If NASA wants to increase diversity, equity, inclusion, and accessibility in the leadership of competed space missions, a very obvious way to do that would be to fund more small satellites. For the cost of a single SMEX (\$150M), $\geq$20 new leaders\footnote{20 Principal Investigators + any other leaders such as Project Scientists, Project Managers, and/or Systems Engineers} could be minted via small satellites (\$7.5M average cost). That much opportunity makes it much easier to ensure that DEI goals are met -- and met soon -- compared to a single SMEX every 2--3 years. We reiterate that all mission classes are needed for science, but small satellites present a clear opportunity to meet NASA's DEI goals. We also note that these benefits can extend further down the pipeline: the low-cost/higher-risk stance makes it possible for projects to engage with high school students, exposing them to Science, Technology, Engineering, and Math (STEM) in a much deeper way that may not have been an opportunity otherwise. The MinXSS CubeSat (funded by NASA, led by University of Colorado), for example, involved over 40 students spanning graduate to high school. This inspiration acts as a force to help individuals from diverse backgrounds stick with STEM by increasing their motivation and providing an exceptional experience they can cite in, e.g., college applications and interviews for research positions. 

Moreover, there should not be a bottleneck in the pipeline just before the opportunity to lead parts of, or entire, large missions becomes available. The current funding level is a bottleneck, with only 1--2 selections possible per year. If NASA wants a more diverse set of leaders, not just of small missions but for large ones, there should be a large pool of exceptionally qualified candidates made available. One highly efficient way for those candidates to gain that qualification is to have received the kind of career development provided by small satellites, as described in Section \ref{sec:career}. We also emphasize that even if a mission never obtains its science measurements, \textbf{the mission can still be a full success on the career development and equity dimensions.}

\section{\textit{How} SmallSats Achieve This High ROI}
\label{sec:how}
The crux of the method for achieving this high ROI is a result of what is illustrated in Figure \ref{fig:ecosystem}: the higher the mission classification, the less technical risk tolerance there is. Money has to be spent to buy down risk. A dimension not shown in the figure is mission scope. While it is true that the mission classifications and scope are correlated, it is possible to imagine a CubeSat mission developed with Class A requirements. The scope wouldn't increase but a great deal of money would have to be spent to buy down the technical risks. An important part of the high ROI is simply that the culture is more risk tolerant with these smaller missions. The relationship between dollars spent specifically to buy down risk and probability of mission success is nonlinear. That is, it costs a lot less money to increase the probability of success from 50\% to 80\% than it does to get from 99.5\% to 99.9\%. As a result, the cost-normalized productivity of small satellites is very high.

The higher risk tolerance also ties into career development: it is more acceptable for mistakes to occur due to lack of experience. The stakes are lower. Those mistakes represent learning, training, and the \textit{gaining} of experience -- in other words, career development.

Finally, investing in small satellites would mean increasing the number of opportunities. The return on that investment manifests in career development, as above, as well as meeting DEI goals. In the limiting case, we could fund everyone who had the desire to lead a mission and no systematic bias would skew the demographics. As that pipeline narrows towards the opposite limiting case, we reach the current state of affairs that has prompted all of the reports and calls for improving DEI in spaceflight leadership. 

\section{Recommendations}
\label{sec:recommendations}

\begin{enumerate}
    \item Funding agencies should give small satellite funding high priority. The number of heliophysics small satellite selections made each year should be increased. The selection ratio in all related program elements should be above 15\%. NASA, in particular, should target a selection ratio commensurate with the average of all ROSES proposals: 30\%\footnote{The \href{https://science.nasa.gov/researchers/sara/grant-stats}{list of 1180 ROSES selections made since 2003} has mean = 32\%, median = 28\% selection ratios}.

    \item NASA's Heliophysics Division should add a Pioneers class (\$20M cost cap), following the example of Astrophysics. This would fill in an important gap between the current average CubeSat cost (\$7.5M) and the Missions of Opportunity (\$50M; see Figure \ref{fig:ecosystem}).
    
    \item NASA's Heliophysics Division should provide more funding for the Heliophysics Flight Opportunities Studies (HFOS) program element. This program funds concept study development with the express purpose to generate more competitive submissions to HFORT. This program is relatively new, so this recommendation is contingent on NASA finding that the program has been successful. Early indicators are positive, though the cost cap per investigation is restrictive. \textbf{Programs like HFOS are especially important for ensuring equity} because they provide resources for teams that happen to be led from institutions that don't have the internal resources necessary to build a compelling mission proposal.
    
    \item NASA's Heliophysics Division should continue the excellent support of instrument development in the Heliophysics Instrument Development for Science (HTIDeS). Special attention to instrument miniaturization in the call for proposals would further incentivize development that can be leveraged by HFORT.
    
    \item Funding agencies should periodically review where their small satellite programs fall and which way they are trending in the ecosystem illustration (Figure \ref{fig:ecosystem}). If the trends are pushing small satellites toward being too risk averse, policies should be amended (e.g., a more liberal risk posture, allowing more tailoring of Class D specification) to ensure that these crucial platforms fill the important niche they presently occupy. 
    
    \item NASA should continue to bolster the Small Satellite and Special Projects Office (S3PO) and Small Spacecraft Systems Virtual Institute (S3VI) to support the growing community. The S3VI seminar series and the S3PO communication of lessons learned across projects are both examples of excellent information dissemination. However, more could be done to facilitate interactions with in the community itself. The first \href{https://sites.wff.nasa.gov/code810/Symposium/Sounding-Rocket-Symposium.html}{Sounding Rocket Symposium} will take place this year; a similar workshop for SmallSats would be welcome. A SmallSat Academy could be established, possibly facilitated through a CubeSat Center of Excellence, to enable the ``see one, do one, teach one" approach described in Section \ref{sec:career}.
\end{enumerate}

\bibliography{references.bib}

\end{document}